\documentclass[aps,pre,floatfix]{revtex4}
\usepackage{epsf}
\usepackage{subfigure}
\usepackage{amsmath}
\usepackage{amssymb}
\usepackage{graphicx}
\usepackage{epstopdf}
\begin{document}

\newcommand{\be}{\begin{equation}}
\newcommand{\ee}{\end{equation}}
\newcommand{\bea}{\begin{eqnarray}}
\newcommand{\eea}{\end{eqnarray}}
\newcommand{\mean}[1]{\left \langle #1 \right \rangle}

\title{\bf Fluctuation relations for equilibrium states with broken discrete symmetries}

\author{Pierre Gaspard}
\affiliation{Center for Nonlinear Phenomena and Complex Systems \& Department of Physics,\\
Universit\'e Libre de Bruxelles, Code Postal 231, Campus Plaine, 
B-1050 Brussels, Belgium}

\begin{abstract}
Relationships are obtained expressing the breaking of spin-reversal symmetry by an external magnetic field in Gibbsian canonical equilibrium states of spin systems under specific assumptions.  These relationships include an exact fluctuation relation for the probability distribution of the magnetization, as well as a relation between the standard thermodynamic entropy, an associated spin-reversed entropy or coentropy, and the product of the average magnetization with the external field, as a non-negative Kullback-Leibler divergence.  These symmetry relations are applied to the model of noninteracting spins, the 1D and 2D Ising models, and the Curie-Weiss model, all in an external magnetic field.  The results are drawn by analogy with similar relations obtained in the context of nonequilibrium physics.

\vskip 0.3 cm

\end{abstract}

\noindent {}

\vskip 0.5 cm

\maketitle

\section{Introduction}
\label{Intro}

The breaking of a discrete symmetry by an external field is a common phenomenon in condensed matter at equilibrium.  Examples are given by spin systems in a uniform external magnetic field, such as the Ising or Curie-Weiss models.  In the presence of the external field, the Hamiltonian of these systems is mapped onto the Hamiltonian in the opposite field under spin reversal.  Concomitantly, the system acquires a magnetization, which breaks the spin-reversal symmetry of the zero-field Hamiltonian.  In finite systems or in finite regions of infinite systems at equilibrium, the order parameter given by the magnetization fluctuates around its average value according to some probability distribution and we may wonder if this probability distribution obeys general relationships resulting from the underlying spin-reversal symmetry.

The similarity is striking with the situation in nonequilibrium open systems in contact with reservoirs where the thermodynamic forces induce currents breaking the time-reversal symmetry.  Nevertheless, the current fluctuations obey remarkable symmetry relations known under the name of fluctuation theorems \cite{ECM93,GC95,K98,LS99,M99,C99,J00,AG07JSP}.  The question arises whether similar results can be obtained for the fluctuations of the order parameter in equilibrium systems where discrete symmetries such as the spin-reversal symmetry are broken.

The purpose of the present paper is to show that it is indeed the case for equilibrium systems under specific assumptions.  We consider spin systems described by Gibbsian equilibrium canonical distributions.  The Hamiltonian is supposed to depend linearly on the order parameter multiplied by the external field, as it is the case in the Ising and Curie-Weiss models.  The fluctuations of the order parameter are characterized in terms of the equilibrium probability distribution of the magnetization and associated large-deviation functions such as the cumulant generating function of the magnetization.  Using the aforementioned symmetry of the Hamiltonian, a fluctuation relation is here proved in a form that is similar to those previously obtained for nonequilibrium steady states. 

Furthermore, a spin-reversed entropy or coentropy is introduced besides the standard thermodynamic entropy in order to characterize the probability of spin configurations that are opposite to the typical configurations. The spin-reversed entropy is analogous to the time-reversed entropy that has been previously introduced in the context of nonequilibrium statistical mechanics  \cite{G04JSP}.  Here, the difference between the spin-reversed and the standard entropies turns out to define a Kullback-Leibler divergence that is always non-negative and proportional to the external field multiplied by the average magnetization.  This result completes the analogy with the time-reversal symmetry breaking by nonequilibrium steady states, for which similar relationships have been established \cite{AG07JSP,G04JSP,G07CRP}.  The theory is illustrated with several systems: (1) a system of noninteracting spins in an external magnetic field; (2) the one-dimensional Ising model; (3) the Curie-Weiss model; (4) the two-dimensional Ising model.
The present paper extensively develops preliminary results reported in Ref.~\cite{G11}.

The paper is organized as follows.  The framework is presented in Section \ref{Framework}.  The equilibrium fluctuation relation is proved in Section \ref{SR}.  In Section \ref{R-entropy}, the spin-reversed entropy is introduced and shown to combine with the standard entropy to get the non-negativity of the average magnetization multiplied by the external magnetic field.  The theory is applied to the system of noninteracting spins in an external magnetic field in Section \ref{Free_spins}, to the one-dimensional Ising model in Section \ref{1DIsing}, to the Curie-Weiss model in Section \ref{Curie_Weiss}, and to the two-dimensional Ising model in Section \ref{2DIsing}.  Conclusions are drawn in Section \ref{Conclusions}.

\section{Spin-reversal symmetry and its breaking by an external magnetic field}
\label{Framework}

\subsection{The Hamiltonian and its symmetries}

We consider systems composed of $N$ spins $\pmb{\sigma}=\{\sigma_i\}_{i=1}^N$.  For spin one-half, the individual spin variables take the values $\sigma_i=\pm 1$ so that the state space $\Sigma=\{\pmb{\sigma}\}$ contains $2^N$ spin configurations.  For spin one, the values would be $\sigma_i=+1,0,-1$ and the state space would contain $3^N$ spin configurations. 

The energy of the system is given by the Hamiltonian function $H_N(\pmb{\sigma};B)$ where $B$ denotes the external magnetic field.  We introduce the magnetization
\be
M_N(\pmb{\sigma})=\sum_{i=1}^N \sigma_i
\label{M}
\ee
which plays the role of order parameter.
The Hamiltonian is assumed to depend linearly on the external magnetic field multiplied by the magnetization
\be
H_N(\pmb{\sigma};B) = H_N(\pmb{\sigma};0)- B \, M_N(\pmb{\sigma})
\label{H}
\ee
as it is the case for instance in the Ising or Curie-Weiss models.

We consider the discrete symmetry of {\it spin reversal}:
\be
\pmb{\sigma}^{\rm R}=R\, \pmb{\sigma}=-\pmb{\sigma}
\label{spin-rev}
\ee
which is an involution because $R^2=1$. Accordingly, the spin-reversal transformation generates the discrete group ${\mathbb Z}_2=\{ 1,R\}$.  We notice that the set of all the spin configurations $\{\pmb{\sigma}\}$ that defines the state space is mapped onto itself by spin reversal, $R\Sigma=R\{\pmb{\sigma}\}=\{\pmb{-\sigma}\}=\{\pmb{\sigma}\}=\Sigma$, because the state space contains the reversal of every spin configuration.

In the absence of external magnetic field, the Hamiltonian is supposed to be symmetric under spin reversal while the magnetization is reversed:
\bea
R \, H_N(\pmb{\sigma};0) \, R &=& H_N(\pmb{\sigma};0)  \label{H-sym}\\
R \, M_N(\pmb{\sigma}) \, R &=& - M_N(\pmb{\sigma})\label{M-sym}
\eea
As a consequence, spin reversal maps the Hamiltonian function in the external magnetic field $B$ onto the Hamiltonian in the reversed magnetic field $-B$:
\be
R \, H_N(\pmb{\sigma};B) \, R = H_N(\pmb{\sigma};-B) 
\label{H-B}
\ee
Therefore, the external field is expected to induce a magnetization, which breaks the ${\mathbb Z}_2$ symmetry of the Hamiltonian in the absence of external field.

Moreover, the Hamiltonian $H_N(\pmb{\sigma};0)$ may also be symmetric under the transformations of another group $G$ of transformations.  This group depends on the geometry of the network of interactions between the spins.  If the spins interact between the nearest neighbors of a lattice on a torus, the group $G$ contains all the translations of the lattice, as in Ising models.  If all the spins interact together in a fully connected graph, the group $G$ is composed of all the $N!$ permutations of the vertices of the graph: $G={\rm Sym}\,N$, as in the Curie-Weiss model.

\subsection{Gibbs' canonical equilibrium states and thermodynamics}

The system is supposed to be in equilibrium at the temperature $T$, in which case the invariant probability distribution is given by Gibbs' canonical equilibrium state:
\be
\mu_N(\pmb{\sigma};B) = \frac{1}{Z_N(B)} \, {\rm e}^{-\beta H_N(\pmb{\sigma};B)} 
\label{Gibbs}
\ee
where $\beta=(kT)^{-1}$ is the inverse temperature and $k$ Boltzmann's constant.  The normalization of the probability distribution $\sum_{\pmb{\sigma}} \mu_N(\pmb{\sigma};B)=1$ is guaranteed by the partition function
\be
Z_N(B) = {\rm tr}\, {\rm e}^{-\beta H_N(\pmb{\sigma};B)}  = \sum_{\pmb{\sigma}} {\rm e}^{-\beta H_N(\pmb{\sigma};B)} 
\ee
Because of the symmetry (\ref{H-B}), the partition function is an even function of the magnetic field:
\be
Z_N(B)=Z_N(-B)
\label{Z-sym}
\ee
since summing over all the spin configurations is equivalent to summing over all their reversals: $\sum_{\pmb{\sigma}}(\cdot)=\sum_{-\pmb{\sigma}}(\cdot)=\sum_{R\pmb{\sigma}}(\cdot)$.

The Helmholtz free energy is defined by
\be
F_N(B) = -kT \, \ln Z_N(B)
\label{F}
\ee
The energy is given by the statistical average of the Hamiltonian function
\be
E_N = \langle H_N\rangle_B = {\rm tr}\, \mu_N \, H_N
\label{E}
\ee
with respect to the Gibbsian probability distribution (\ref{Gibbs}),
while the entropy is defined by its thermodynamic relation:
\be
S_N = - \frac{\partial F_N}{\partial T}
\label{S}
\ee
so that the basic relation defining the Helmholtz free energy is satisfied: $F_N=E_N-TS_N$.

The statistical average of the total magnetization is obtained by taking the derivative of the free energy with respect to the external field:
\be
\langle M_N\rangle_B = \frac{\partial F_N}{\partial B}
\ee
The free energy per spin is defined by
\be
f(B) = \lim_{N\to\infty} \frac{1}{N} \, F_N(B)
\ee
and the average magnetization per spin by
\be
\langle m\rangle_B = \lim_{N\to\infty} \frac{1}{N} \, \langle M_N\rangle_B = - \frac{\partial f}{\partial B}
\label{m_spin}
\ee

Now, the Legendre transform of the free energy may be introduced as
\be
g(m) = f(B) + m\, B
\label{g}
\ee
where $B=B(m)$ is the magnetic field corresponding to the given magnetization $m=-\partial_Bf[B(m)]$.
Reciprocally, the free energy per spin can be recovered by the Legendre transform of the new function $g(m)$, in which the magnetization per spin is replaced by its value corresponding to the external field $B=\partial_mg[m(B)]$.  In systems with phase transitions where these derivatives may not exist, the Legendre transform should be replaced by its extension called the Legendre-Fenchel transform \cite{E85,T09}.

We notice that the symmetry (\ref{Z-sym}) of the partition function results into the symmetry relations:
\be
f(B)=f(-B) \qquad\qquad\mbox{and}\qquad\qquad g(m)=g(-m)
\label{f-g-sym}
\ee

Furthermore, the magnetic susceptibility per spin is defined as
\be
\chi_B = \lim_{N\to\infty} \frac{1}{N} \, \frac{\partial \langle M_N\rangle_B}{\partial B}
\ee
and the variance of the magnetization fluctuations as
\be
\sigma_B^2  = \lim_{N\to\infty} \frac{1}{N} \, \left(\langle M_N^2\rangle_B - \langle M_N\rangle_B^2\right)
\ee

\section{Symmetry relations}
\label{SR}

\subsection{The fluctuation relation}

We introduce the probability $P_B(M)$ that the fluctuating magnetization would take the value $M=M_N(\pmb{\sigma})$ as
\be
P_B(M) =\langle \delta_{M,M_N(\pmb{\sigma})}\rangle_B
\label{P}
\ee
where $\delta_{M,M'}$ denotes the Kronecker delta function defined for $M,M'\in{\mathbb Z}$ and such that $\delta_{M,M'}=1$ if $M=M'$ and zero otherwise.  This probability distribution is normalized to unity according to $\sum_{M}P_B(M)=1$.  The statistical average is carried out over the Gibbsian canonical equilibrium distribution (\ref{Gibbs}).

The fluctuation relation is established as follows:
\bea
P_B(M) &=& \frac{1}{Z_N(B)} \sum_{\pmb{\sigma}} {\rm e}^{-\beta H_N(\pmb{\sigma};0)+\beta B M_N(\pmb{\sigma})} \; \delta_{M,M_N(\pmb{\sigma})} \\
&=& 
\frac{1}{Z_N(B)} \sum_{\pmb{\sigma}^{\rm R}} {\rm e}^{-\beta H_N(\pmb{\sigma}^{\rm R};0)+\beta B M_N(\pmb{\sigma}^{\rm R})} \; \delta_{M,M_N(\pmb{\sigma}^{\rm R})} \\
&=&
\frac{1}{Z_N(B)} \sum_{\pmb{\sigma}} {\rm e}^{-\beta H_N(\pmb{\sigma};0)-\beta BM_N(\pmb{\sigma})} \; \delta_{M,-M_N(\pmb{\sigma})} \\
&=&
\frac{1}{Z_N(B)}\; {\rm e}^{2\beta B M} \sum_{\pmb{\sigma}} {\rm e}^{-\beta H_N(\pmb{\sigma};0)+\beta BM_N(\pmb{\sigma})} \; \delta_{-M,M_N(\pmb{\sigma})} \\
&=& {\rm e}^{2\beta B M} \; P_B(-M)
\eea
From the first to the second line, the sum over all the spin configurations is equivalent to the sum over the spin-reversed configurations since they belong to the same set: $R\Sigma=\Sigma$.  From the second to the third line, the symmetries (\ref{H-sym}) and (\ref{M-sym}) are used so that $H_N(\pmb{\sigma}^{\rm R};0)=H_N(\pmb{\sigma};0)$ and $M_N(\pmb{\sigma}^{\rm R})=-M_N(\pmb{\sigma})$.  From the third to the fourth line, the Kronecker delta function allows us to restore the canonical equilibrium distribution by factorizing $\exp(2\beta B M)$ out of the sum. Finally, we obtain the fluctuation relation:
\be
\frac{P_B(M)}{P_B(-M)} = {\rm e}^{2\beta B M} 
\label{FR}
\ee
as an exact relationship under the aforementioned assumptions.  This relation is one of the main results of this paper.  We notice the analogy with the well-known fluctuation theorems for nonequilibrium steady states \cite{ECM93,GC95,K98,LS99,M99,C99,J00,AG07JSP}.  This analogy shows that these symmetry relations express the breaking of the discrete symmetry by the probability distribution in the presence of external constraints, here given by the magnetic field $B$.  Indeed, in the absence of the magnetic field $B=0$, the opposite fluctuations of the magnetization have equal probabilities and the symmetry is thus recovered.  If the external field is non vanishing, a bias appears between opposite fluctuations and the fluctuations become more probable in one particular direction, which nicely expresses the breaking of the symmetry by the external field.

\subsection{The large-deviation function}

In the large-system limit, the probability distribution of the total magnetization is expected to behave exponentially in the number $N$ of spins.  In this regard, we can introduce the large-deviation function
\be
\Phi_B(m) = \lim_{N\to\infty} -\frac{1}{N}\, \ln P_B(Nm)
\label{Phi}
\ee
Therefore, the probability distribution (\ref{P}) should behave as
\be
P_B(M) \sim {\rm e}^{-N\Phi_B(M/N)} \qquad\mbox{for}\qquad N\to\infty
\label{P-Phi}
\ee
up to a subexponential dependence on $N$ \cite{E85,T09,TH12}.

Now, the fluctuation relation (\ref{FR}) implies that the large-deviation function (\ref{Phi}) obeys the following symmetry relation:
\be
\Phi_B(-m)-\Phi_B(m) = 2\beta B m
\label{Phi-sym}
\ee
as can be verified by a straightforward calculation.

The large-deviation function can be expressed in terms of the Helmholtz free energy per spin and its Legendre transform (\ref{g}) according to
\be
\Phi_B(m) = \beta \left[ g(m) - B\, m - f(B)\right]
\label{Phi-g-f}
\ee
In order to obtain this result, we notice that the partition function can be written as
\be
Z_N(B) = \sum_M {\rm e}^{\beta B M} \sum_{\pmb{\sigma}} {\rm e}^{-\beta H_N(\pmb{\sigma};0)} \; \delta_{M,M_N(\pmb{\sigma})}
\ee
by introducing the Kronecker delta function.
Since the partition function behaves as $Z_N(B)\simeq{\rm e}^{-N\beta f(B)}$ in the large-system limit, we see that the sum over the spin configurations should behave as
\be
\sum_{\pmb{\sigma}} {\rm e}^{-\beta H_N(\pmb{\sigma};0)} \; \delta_{Nm,M_N(\pmb{\sigma})} \sim {\rm e}^{-N\beta g(m)} \qquad\mbox{for}\qquad N\to\infty
\ee
in order to recover the function (\ref{g}) as the Legendre transform of the free energy per spin $f(B)$.
Since the probability distribution of the magnetization can be obtained from the expression
\be
P_B(M) = \frac{1}{Z_N(B)} \; {\rm e}^{\beta B M} \sum_{\pmb{\sigma}} {\rm e}^{-\beta H_N(\pmb{\sigma};0)} \; \delta_{M,M_N(\pmb{\sigma})}
\ee
we find that the large-deviation function is indeed given by  Eq.~(\ref{Phi-g-f}) in the large-system limit.

We notice that the symmetry relation (\ref{Phi-sym}) is verified because of the symmetry (\ref{f-g-sym}) of the function $g(m)$.

\subsection{The cumulant generating function and its symmetry}

In the framework of large-deviation theory  \cite{T09,TH12}, the cumulant generating function of the magnetization fluctuations is defined as
\be
Q_B(\lambda) = \lim_{N\to\infty} -\frac{1}{N} \, \ln \langle {\rm e}^{-\lambda M_N}\rangle_B
\label{Q}
\ee
This function is generating in the sense that the average magnetization per spin, its variance, as well as all the higher-order cumulants can be obtained by taking successive derivatives with respect to the parameter $\lambda$:
\be
\langle m\rangle_B = \frac{\partial Q_B}{\partial \lambda}(0)\; , \qquad \sigma_B^2 = -\frac{\partial^2Q_B}{\partial \lambda^2}(0) \; , \qquad ...
\ee

Now, the statistical average in Eq.~(\ref{Q}) can be written in terms of the probability distribution (\ref{P}) of the magnetization as
\be
\langle {\rm e}^{-\lambda M_N}\rangle_B = \sum_M P_B(M) \, {\rm e}^{-\lambda M}
\label{aver-exp-P}
\ee
Because of the fluctuation relation (\ref{FR}), we immediately obtain the symmetry relation of the cumulant generating function:
\be
Q_B(\lambda) = Q_B(2\beta B-\lambda)
\label{Q-sym}
\ee
which is similar to symmetry relations obtained elsewhere for nonequilibrium systems \cite{AG07JSP}.

In analogy with the results of Ref.~\cite{AG07JSM}, the response coefficients can be related to the cumulants of the magnetization fluctuations be taking successive derivatives of the symmetry relation (\ref{Q-sym}) with respect to the parameter $\lambda$ and the external field $B$.  In particular, taking one derivative $\partial_\lambda$ and another $\partial_B$ gives
\be
\frac{\partial^2Q_B}{\partial\lambda\partial B}(\lambda) = - \frac{\partial^2Q_B}{\partial\lambda\partial B}(2\beta B-\lambda)  -2\, \beta\, \frac{\partial^2Q_B}{\partial\lambda^2}(2\beta B-\lambda) 
\ee
and setting $\lambda=B=0$, we recover the well-known relation
\be
\chi_0 = \beta \, \sigma_0^2
\label{chi-sigma}
\ee
between the magnetic susceptibility and the variance of the magnetization fluctuations in the absence of external field.  Similar relationships can be obtained between higher-order response coefficients and cumulants.  We notice that, for Gibbsian equilibrium states, the relation (\ref{chi-sigma}) also holds for non-vanishing external fields, $\chi_B = \beta\, \sigma_B^2$, which is not the case in nonequilibrium steady states since these latter do not have the simple Gibbsian form (\ref{Gibbs}).

The cumulant generating function can also be obtained in terms of the partition function and the free energy per spin.  Indeed, we have that
\be
\langle {\rm e}^{-\lambda M_N}\rangle_B = \sum_{\pmb{\sigma}} \mu_N(\pmb{\sigma};B) \, {\rm e}^{-\lambda M_N(\pmb{\sigma})} = \frac{1}{Z_N(B)}\sum_{\pmb{\sigma}} {\rm e}^{-\beta H_N(\pmb{\sigma};0)+(\beta B-\lambda) M_N(\pmb{\sigma})} = \frac{Z_N(B-\beta^{-1}\lambda)}{Z_N(B)}
\label{aver-exp}
\ee
In the thermodynamic limit $N\to\infty$, we thus obtain
\be
Q_B(\lambda) = \beta \left[ f(B-\beta^{-1}\lambda) - f(B)\right]
\ee
In this regard, the symmetry (\ref{Z-sym}) for the partition function or (\ref{f-g-sym}) for the free energy per spin implies the symmetry relation (\ref{Q-sym}) for the cumulant generating function.

A further consequence of Eq.~(\ref{aver-exp-P}) is that the cumulant generating function (\ref{Q}) can be obtained as the Legendre transform of the large-deviation function (\ref{Phi}):
\be
Q_B(\lambda) =  \Phi_B[m(\lambda)] + \lambda \, m(\lambda) \qquad\mbox{where}\qquad
\frac{\partial \Phi_B}{\partial m}[m(\lambda)] = - \lambda
\ee
Conversely, the large-deviation function is given by the Legendre transform of the cumulant generating function:
\be
\Phi_B(m) = Q_B[\lambda(m)] - m \, \lambda(m)
\qquad\mbox{where} \qquad \frac{\partial Q_B}{\partial\lambda}[\lambda(m)] =m
\label{Legendre-Q>Phi}
\ee
The Legendre-Fenchel transform should be used if these functions were non differentiable \cite{E85,T09,TH12}.

\section{Entropy, coentropy and broken symmetry}
\label{R-entropy}

In this section, we complete the parallelism with the results obtained for nonequilibrium steady states
by considering  a spin-reversed entropy or coentropy defined in analogy with the time-reversed entropy per unit time that has been previously introduced \cite{G04JSP}.  This quantity is shown to combine with the standard thermodynamic entropy to form a Kullback-Leibler divergence, as well as an important relationship with the average value of the magnetization.

\subsection{Entropy and coentropy}

The standard thermodynamic entropy is defined by Eq.~(\ref{S}), from which we get the usual expression:
\be
S_N = -k \, {\rm tr}\, \mu_N \ln \mu_N
\ee
for the Gibbsian equilibrium state (\ref{Gibbs}).

The breaking of the spin-reversal symmetry by the external field can be characterized by comparing the probability $\mu_N(\pmb{\sigma};B)$ of some spin configuration $\pmb{\sigma}$ with the probability of the reversed configuration:
\be
\mu_N^{\rm R}(\pmb{\sigma};B) = \mu_N(R\pmb{\sigma};B) = \mu_N(-\pmb{\sigma};B)
\ee
Here, we introduce the {\it spin-reversed entropy}
\be
S_N^{\rm R} = -k \, {\rm tr}\, \mu_N  \ln \mu_N^{\rm R}
\ee
in analogy with a similar quantity introduced elsewhere in the context of nonequilibrium statistical mechanics \cite{G04JSP}.

Now, the difference between the spin-reversed and the standard entropy defines the Kullback-Leibler divergence:
\be
D(\mu_N \Vert \mu_N^{\rm R}) = {\rm tr}\, \mu_N  \ln \frac{\mu_N}{\mu_N^{\rm R}} = \frac{1}{k} \left( S_N^{\rm R}-S_N\right) \geq 0
\ee
which is known to be always non-negative \cite{CT06}.  In this regard, the spin-reversed entropy could be called a {\it coentropy} since it combines with the entropy to form the non-negative Kullback-Leibler divergence.

The Kullback-Leibler divergence vanishes in the absence of external field when the probabilities of every spin configuration and its reversal are equal $\mu_N=\mu_N^{\rm R}$.  However, in the presence of an external field, the probability of a reversed configuration $R\pmb{\sigma}$ is expected to take a different value than the probability of the configuration $\pmb{\sigma}$ itself, in which case the Kullback-Leibler divergence becomes positive.  Accordingly, the Kullback-Leibler divergence is a measure of the breaking of the symmetry at the level of the probability distribution $\mu_N$.  This is confirmed by the remarkable relation
\be
D(\mu_N \Vert \mu_N^{\rm R}) = \frac{1}{k} \, \left( S_N^{\rm R}-S_N\right) = 2  \beta  B \, \langle M_N\rangle_B \geq 0
\label{D-B-M}
\ee
which is the consequence of the fact that
\be
{\rm tr}\, \mu_N  \ln \frac{\mu_N}{\mu_N^{\rm R}} = 2 \beta  B \; {\rm tr}\, \mu_N   M_N
\ee
holds for the Gibbsian equilibrium state (\ref{Gibbs}).
A corollary of this result is that spin systems described by a Hamiltonian function (\ref{H}) with the symmetries (\ref{H-sym}) and (\ref{M-sym}) is always paramagnetic because the average value of the magnetization should point in the same direction as the external magnetic field.

If we introduce the entropy and coentropy per spin as
\bea
s &=& \lim_{N\to\infty} \frac{1}{N} \, S_N \\
s^{\rm R} &=& \lim_{N\to\infty} \frac{1}{N} \, S_N^{\rm R}
\eea
the relation (\ref{D-B-M}) writes
\be
\frac{1}{k}\left(s^{\rm R} - s\right) = 2\beta B \, \langle m\rangle_B \geq 0
\label{sR-s-Bm}
\ee
where $\langle m\rangle_B$ is the average magnetization per spin (\ref{m_spin}).
Consequently, we have the general inequalities:
\be
s^{\rm R} \geq s \geq 0
\ee
The relation (\ref{sR-s-Bm}) is another of the main results of this paper.

\subsection{The disorders of the spin configurations and their reversals}

In order to interpret the previous result, we consider spin systems on an infinite $d$-dimensional lattice ${\mathbb Z}^d$.  In the infinite-system limit, the equilibrium states 
\be
\mu = \lim_{N\to\infty} \mu_N
\ee
can be constructed as the so-called Dobrushin-Lanford-Ruelle states \cite{E85,D68,LR69}.
Let $\Lambda\subset{\mathbb Z}^d$ be a part of the lattice and $\vert\Lambda\vert$ its volume.  If $\pmb{\sigma}\in \Sigma_{{\mathbb Z}^d}$ is a spin configuration of the infinite lattice, $\pmb{\sigma}_{\Lambda}$ is the spin configuration restricted on the domain $\Lambda$ of the lattice.  
In this framework, the entropy and coentropy per spin are defined as
\bea
s &=& \lim_{\vert\Lambda\vert\to\infty} -\frac{k}{\vert\Lambda\vert} \, {\rm tr}\, \mu(\pmb{\sigma}_{\Lambda})\ln\mu(\pmb{\sigma}_{\Lambda}) \label{s_spin}\\
s^{\rm R} &=& \lim_{\vert\Lambda\vert\to\infty} -\frac{k}{\vert\Lambda\vert} \, {\rm tr}\, \mu(\pmb{\sigma}_{\Lambda})\ln\mu(\pmb{\sigma}_{\Lambda}^{\rm R})
\label{sR_spin}
\eea

These quantities characterize the disorders in the spin configurations and their reversals.  Indeed, for almost every spin configuration $\pmb{\sigma}$ with respect to the probability measure $\mu$, the probabilities that the spins in the domain $\Lambda$ are found in the configuration $\pmb{\sigma}_{\Lambda}$ and their reversals in the configuration $\pmb{\sigma}_{\Lambda}^{\rm R}$ should decay in the infinite-volume limit $\vert\Lambda\vert\to\infty$ as
\bea
&& \mu(\pmb{\sigma}_{\Lambda}) \sim {\rm e}^{-\vert\Lambda\vert\, s/k} \\
&& \mu(\pmb{\sigma}_{\Lambda}^{\rm R}) \sim {\rm e}^{-\vert\Lambda\vert\, s^{\rm R}/k} 
\eea
The faster the decay, the rarer the spin configuration $\pmb{\sigma}$ or its reversal $\pmb{\sigma}^{\rm R}$ among all the possible configurations.  

Because of Eq.~(\ref{sR-s-Bm}), we have the general relationship
\be
s^{\rm R} = s + \frac{2}{T} \, B \, \langle m\rangle_B \qquad \mbox{with}\qquad B\, \langle m\rangle_B \geq 0
\label{coentropy}
\ee
between the entropies (\ref{s_spin})-(\ref{sR_spin}) and the average magnetization per spin $\langle m\rangle_B$.  As a consequence, the ratio of the probabilities of opposite spin configurations behaves as
\be
\frac{\mu(\pmb{\sigma}_{\Lambda})}{\mu(\pmb{\sigma}_{\Lambda}^{\rm R})} \simeq {\rm e}^{2\beta B \langle m\rangle_B \vert\Lambda\vert} 
\label{ratio}
\ee
for $\mu$-almost every configuration $\pmb{\sigma}$ in the limit $\vert\Lambda\vert\to\infty$.
If the external field vanishes $B=0$, the spin configurations and their reversals are thus equiprobable.  However, in the presence of an external field breaking the symmetry, the spin reversal of every typical configuration is always less probable than the typical configuration itself because $s^{\rm R}>s\geq 0$ for $B\neq 0$.  In this case, the spin-reversed configurations appear more disordered than the typical configurations themselves.  The analogy is here complete with similar results obtained for nonequilibrium steady states \cite{G04JSP,G07CRP}.  

We notice that there is an important difference between Eq.~(\ref{ratio}) and the fluctuation relation (\ref{FR}).
Indeed, Eq.~(\ref{ratio}) concerns typical spin configurations $\pmb{\sigma}$ characterized by the average magnetization per spin $\langle m\rangle_B$, although the fluctuation relation (\ref{FR}) depends on the variable magnetization $M$ and not on its average value $\langle M\rangle_B$.  

All these results are illustrated with specific models in the following sections.

\section{The noninteracting spin model}
\label{Free_spins}

\subsection{Hamiltonian and thermodynamics}

In this section, we consider the simple model of $N$ noninteracting spins $\sigma_i=\pm 1$ in an external magnetic field $B$.  The Hamiltonian of this system writes
\be
H_N(\pmb{\sigma};B)=-B\sum_{i=1}^N \sigma_i
\ee
that is symmetric under the group $G={\rm Sym} \, N$ and also under the group ${\mathbb Z}_2=\{1,R\}$ if $B=0$.  This latter symmetry is broken if $B\neq 0$.

For an equilibrium spin system at the temperature $T=(k\beta)^{-1}$, the partition function is given by
\be
Z_N(B)={\rm tr}\, {\rm e}^{-\beta H_N} = \left( 2 \, \cosh \beta B\right)^N
\ee
the free energy by
\be
F_N=- NkT \ln\left( 2 \cosh\beta B\right)
\ee
and the energy by
\be
E_N=\langle H_N\rangle_B = -B \, \langle M_N\rangle_B
\ee
with the average total magnetization
\be
\langle M_N\rangle_B = N \, \tanh \beta B
\ee
The entropy of this system has Shannon's form
\be
S_N = Nk \left( -p_+ \ln p_+ - p_- \ln p_-\right)
\ee
in terms of the probabilities
\be
p_{\pm} = \frac{{\rm e}^{\pm\beta B}}{{\rm e}^{+\beta B}+{\rm e}^{-\beta B}}
\label{p+-}
\ee
that a spin would be up or down.  These probabilities satisfy the normalization condition $p_++p_-=1$.  The average magnetization per spin is equal to $\langle m \rangle_B = p_+-p_-={\rm tanh}\beta B$.

\subsection{Fluctuation relation}

The probability (\ref{P}) that the fluctuating magnetization would take the value $M$ can be exactly  calculated for the noninteracting spin model.  Indeed, this probability is given by the binomial distribution as
\be
P_B(M) = \frac{N!}{N_+! \, N_-!} \; p_+^{N_+} \, p_-^{N_-}
\label{P-NIS-1}
\ee
with the probabilities (\ref{p+-}) and the numbers $N_{\pm}$ of up and down spins.  These numbers are related to the value $M$ of the magnetization according to
\be
N_{\pm} = \frac{1}{2}\left( N\pm M\right)
\ee
so that the probability distribution (\ref{P-NIS-1}) can be written as
\be
P_B(M) = \frac{N!}{\left(\frac{N+M}{2}\right)! \, \left(\frac{N-M}{2}\right)! } \; p_+^{\frac{N+M}{2}} \, p_-^{\frac{N-M}{2}}
\label{P-NIS-2}
\ee
Therefore, we see that the fluctuation relation (\ref{FR}) holds
\be
\frac{P_B(M)}{P_B(-M)} = \left(\frac{p_+}{p_-}\right)^M = {\rm e}^{2\beta B M}
\label{FR-NIS}
\ee
because of Eqs.~(\ref{p+-}).

Here, the cumulant generating function (\ref{Q}) has the expression
\be
Q_B(\lambda) = \ln \frac{\cosh\beta B}{\cosh (\beta B-\lambda)}
\ee
so that the symmetry relation (\ref{Q-sym}) is satisfied.  The  Legendre transform (\ref{Legendre-Q>Phi}) gives the large-deviation function
\be
\Phi_B(m) = \frac{1+m}{2}\, \ln \frac{1+m}{2} + \frac{1-m}{2}\, \ln \frac{1-m}{2} - \beta B m +\ln\left(2\cosh \beta B\right)
\ee
This function determines the asymptotic behavior of the probability distribution (\ref{P-NIS-2}) in the large-system limit $N\to\infty$ and its symmetry relation (\ref{Phi-sym}) is indeed satisfied.

\subsection{Statistics of typical and reversed microstates}

The probability of a microstate or spin configuration $\pmb{\sigma}$ with $N_\pm$ spins that are up or down is given by
\be
\mu_N(\pmb{\sigma}) = p_+^{N_+} \, p_-^{N_-}
\ee
and the probability of the corresponding spin-reversed configuration by
\be
\mu_N(\pmb{\sigma}^{\rm R}) = p_+^{N_-} \, p_-^{N_+}
\ee
The ratio of both probabilities takes the exact value
\be
\frac{\mu_N(\pmb{\sigma})}{\mu_N(\pmb{\sigma}^{\rm R})} = \left(\frac{p_+}{p_-}\right)^{N_+-N_-} = {\rm e}^{2\beta B (N_+-N_-)}
\label{ratio-NIS}
\ee

Now, let us consider these expressions for the most probable configurations, i.e., for the typical microstates such that
\be
N_+\simeq N\, p_+ \qquad\mbox{and}\qquad N_-\simeq N\, p_-
\label{typical}
\ee
in the large-system limit $N\to\infty$.  Using the Stirling formula, the total number of these typical microstates can be evaluated as
\be
\frac{N!}{N_+! \, N_-!} \simeq_{N\to\infty} \frac{1}{\sqrt{2\pi  p_+p_- N}} \; {\rm e}^{Ns/k}
\ee
in terms of the standard thermodynamic entropy per spin:
\be
s=-k\left( p_+\ln p_+ + p_- \ln p_-\right)
\ee
If this entropy is non vanishing, the probability that one among all the typical microstates is observed is thus decaying exponentially as
\be
\mu_N(\pmb{\sigma}) \simeq_{N\to\infty} {\rm e}^{N(p_+\ln p_+ + p_- \ln p_-)} = {\rm e}^{-Ns/k}
\ee
at the rate given by the entropy per spin,
while the probability of the corresponding spin-reversed microstates decays as
\be
\mu_N(\pmb{\sigma}^{\rm R}) \simeq_{N\to\infty} {\rm e}^{N(p_-\ln p_+ + p_+ \ln p_-)} = {\rm e}^{-Ns^{\rm R}/k}
\ee
at the rate 
\be
s^{\rm R}=-k\left( p_-\ln p_+ + p_+ \ln p_-\right)
\ee
that is the coentropy per spin. Here, we can directly verify that the ratio of both probabilities indeed satisfies the relation (\ref{ratio}) because, for $\mu_N$-almost all the microstates $\pmb{\sigma}$, we have that
\be
\frac{\mu_N(\pmb{\sigma})}{\mu_N(\pmb{\sigma}^{\rm R})} \simeq {\rm e}^{N(s^{\rm R}-s)/k}
\label{ratio-almost-NIS}
\ee
with
\be
\frac{1}{k} \, \left( s^{\rm R}-s\right) = \left( p_+-p_-\right) \, \ln \frac{p_+}{p_-} = 2\beta B \tanh\beta B = 2\beta B \langle m \rangle_B \geq 0
\label{KLdist-NIS}
\ee
This quantity is non negative because it is related to the Kullback-Leibler divergence (\ref{D-B-M}).
Another way to verify the result (\ref{ratio}) is to use directly Eq.~(\ref{ratio-NIS}) on the subset of the most probable microstates for which we have that
\be
N_+-N_- \simeq N(p_+-p_-) = N \, \tanh \beta  B = N\, \langle m\rangle_B
\ee
leading to the same result (\ref{ratio-almost-NIS}) with Eq.~(\ref{KLdist-NIS}).

\begin{figure}[h]
\begin{center}
\includegraphics[scale=0.45]{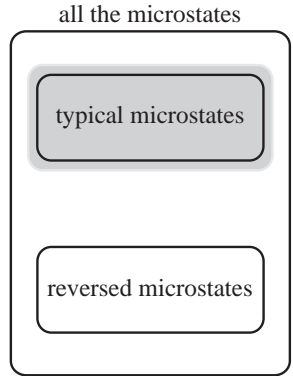}
\caption{Schematic representation of the set of all the $2^N$ possible microstates containing the subset of the typical microstates with $N_+=Np_+$ spins up and $N_-=Np_-$ spins down together with the subset of the corresponding spin-reversed microstates with $N_-=Np_-$ spins up and $N_+=Np_+$ spins down.  The grey area indicates the region where the probability distribution is concentrated.  The probability distribution culminates on the typical microstates that are the most probable microstates. The subsets of typical and reversed microstates coincide if $N_+=N_-$, but they do not overlap as soon as $N_+\neq N_-$.}
\label{fig1}
\end{center}
\end{figure}

In the case of a non-vanishing external field $B\neq 0$ breaking the ${\mathbb Z}_2$ symmetry, the situation is illustrated in Fig.~\ref{fig1}, which depicts the set of all the possible microstates of the spin system.  On this set, the Gibbsian equilibrium probability distribution is concentrated in a certain area shown in grey where we find the subset of the most probable microstates, called the typical microstates, for which Eqs.~(\ref{typical}) hold.  Because $N_+\neq N_-$ if $B\neq 0$, the corresponding spin-reversed microstates form a disjoint subset in this case.  The spin reversals of the typical microstates are thus found in an area of low probability because of the general relation (\ref{ratio-almost-NIS}).  

In the model of noninteracting spins, the validity of the fluctuation relation and the other symmetry relations can thus be verified by direct calculations.

\section{The 1D Ising model}
\label{1DIsing}

\subsection{Hamiltonian and thermodynamics}

A further model where the symmetry relations can be verified analytically is the 1D Ising model
of Hamiltonian
\be
H_N(\pmb{\sigma};B) = - J \sum_{i=1}^N \sigma_i \; \sigma_{i+1} - B \sum_{i=1}^N \sigma_i
\ee
where the $N$ spins $\sigma_i=\pm 1$ form a ring with $\sigma_{N+1}=\sigma_1$.

It is well known that the canonical partition function of this model can be calculated thanks to the transfer-matrix technique \cite{B07}. Accordingly, all the thermodynamic quantities are exactly calculable for this 1D model.

\subsection{Fluctuation relation}

Here, we are interested in the generating function of the statistical moments of the fluctuating total magnetization~$M_N(\pmb{\sigma})$
\be
\langle{\rm e}^{-\lambda M_N}\rangle_B = \frac{1}{Z_N(B)} \, \sum_{\pmb{\sigma}} \exp\left[ \beta J \sum_{i=1}^N \sigma_i\sigma_{i+1} + (\beta B-\lambda) \sum_{i=1}^N \frac{\sigma_i+\sigma_{i+1}}{2}\right]
\ee
 which can be expressed as
\be
\langle{\rm e}^{-\lambda M_N}\rangle_B = \frac{{\rm tr}\, \hat V_\lambda^N}{{\rm tr}\, \hat V_0^N}
\ee
in terms of the following transfer matrix
\be
\hat V_{\lambda} = 
\left(\begin{array}{cc}
{\rm e}^{\beta J + \beta B-\lambda} & {\rm e}^{-\beta J}\\
{\rm e}^{-\beta J} & {\rm e}^{\beta J - \beta B+\lambda}
\end{array}
\right)
\label{transfer}
\ee

\begin{figure}[h]
\begin{center}
\includegraphics[scale=0.55]{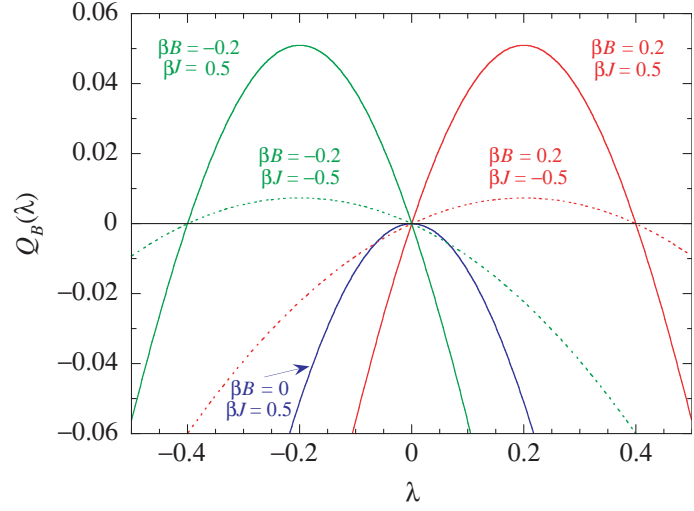}
\caption{Cumulant generating function (\ref{Q-1DIsing}) of the 1D Ising model versus the generating parameter $\lambda$ for $\beta J=0.5$ (solid lines) and $\beta J=-0.5$ (dashed lines).}
\label{fig2}
\end{center}
\end{figure}

This transfer matrix is real and symmetric.  Therefore, it has two real eigenvalues $\Lambda_+(B-\beta^{-1}\lambda) \geq \Lambda_-(B-\beta^{-1}\lambda)$.  The trace of its $N^{\rm th}$ iterate can thus be written as
\be
{\rm tr}\, \hat V_\lambda^N = \Lambda_+(B-\beta^{-1}\lambda)^N + \Lambda_-(B-\beta^{-1}\lambda)^N
\ee
Accordingly, the trace is dominated by the largest eigenvalue $\Lambda_+(B-\beta^{-1}\lambda)$ in the large-system limit $N\to\infty$.  Therefore, the cumulant generating function (\ref{Q}) is obtained as
\be
Q_B(\lambda) = \ln \frac{\Lambda_+(B)}{\Lambda_+(B-\beta^{-1}\lambda)}
\ee
With the explicit expression of the leading eigenvalue $\Lambda_+$, we find that
\be
Q_B(\lambda) = \ln \frac{\cosh\beta B + \sqrt{\sinh^2\beta B + {\rm e}^{-4\beta J}}}{\cosh(\beta B-\lambda) + \sqrt{\sinh^2(\beta B-\lambda) + {\rm e}^{-4\beta J}}}
\label{Q-1DIsing}
\ee
The symmetry relation (\ref{Q-sym}) is thus verified for this model as well.  This symmetry is illustrated in Fig.~\ref{fig2} depicting the cumulant generating function for different values of $\beta J$ and $\beta B$.  In every cases, the function is symmetric under the transformation $\lambda\to 2\beta B-\lambda$.  This is obvious if $B=0$.  If  $\beta B=0.2$ for instance, the symmetry holds under the map $\lambda\to 0.4-\lambda$, as indeed observed.  Another symmetry is also visible in Fig.~\ref{fig2}, namely $Q_B(\lambda)=Q_{-B}(-\lambda)$, which results from the symmetry (\ref{Z-sym}) of the partition function.

\begin{figure}[h]
\begin{center}
\includegraphics[scale=0.55]{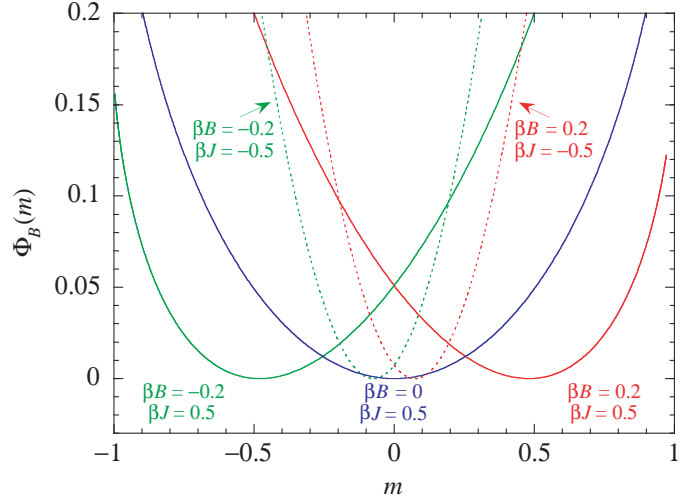}
\caption{The large-deviation function (\ref{LDF-1DIsing}) of the 1D Ising model versus the magnetization variable $m$ for the same values of the physical parameters as in Fig.~\ref{fig2}.}
\label{fig3}
\end{center}
\end{figure}

Moreover, the symmetry $\lambda\to 2\beta B-\lambda$ already manifests itself at the level of the transfer matrix (\ref{transfer}) that satisfies
\be
\hat R \; \hat V_{\lambda} \; \hat R = \hat V_{2\beta B-\lambda} \qquad \mbox{where} \qquad 
\hat R =
\left(\begin{array}{cc}
0 & 1\\
1 & 0
\end{array}
\right)
\ee
is the matrix performing spin reversal: $\sigma_i=\pm 1 \to \sigma_i^{\rm R}=\mp 1$.
Consequently, the symmetry $\lambda \to 2\beta B- \lambda$ holds not only for the cumulant generating function (\ref{Q-1DIsing}) determined by the leading eigenvalue $\Lambda_+$, but also for both eigenvalues $\Lambda_\pm$.

From Eq.~(\ref{Q-1DIsing}), the average magnetization per spin is given by
\be
\langle m\rangle_B= \frac{\partial Q_B}{\partial\lambda}(0) = \frac{\sinh \beta B}{\sqrt{\sinh^2 \beta B + {\rm e}^{-4\beta J}}}
\label{m-1DIsing}
\ee
as expected.  The magnetization behaves smoothly in the plane $(T,B)$ except at the point $(T=0,B=0)$ where the function is discontinuous, although this 1D model does not have a genuine phase transition at a positive temperature.

The Legendre transform of the cumulant generating function (\ref{Q-1DIsing}) takes the following form:
\be
\Phi_B(m) = \ln\frac{\sqrt{1-m^2}\left( \cosh\beta B + \sqrt{\sinh^2\beta B + {\rm e}^{-4\beta J}}\right)}{{\rm e}^{-2\beta J} + \sqrt{m^2\, {\rm e}^{-4\beta J}+1 - m^2}} - m \, \ln\frac{\sqrt{1-m^2}}{m \, {\rm e}^{-2\beta J} + \sqrt{m^2\, {\rm e}^{-4\beta J}+1 - m^2}} - \beta B \, m
\label{LDF-1DIsing}
\ee
This function is depicted in Fig.~\ref{fig3} for the same conditions as in Fig.~\ref{fig2}. 
The analytical expression (\ref{LDF-1DIsing}) is such that the symmetry relation (\ref{Phi-sym}) is indeed satisfied, hence the fluctuation relation (\ref{FR}) holds for the 1D Ising model.

\subsection{Entropy and coentropy per spin}

\begin{figure}[h]
\begin{center}
\includegraphics[scale=0.55]{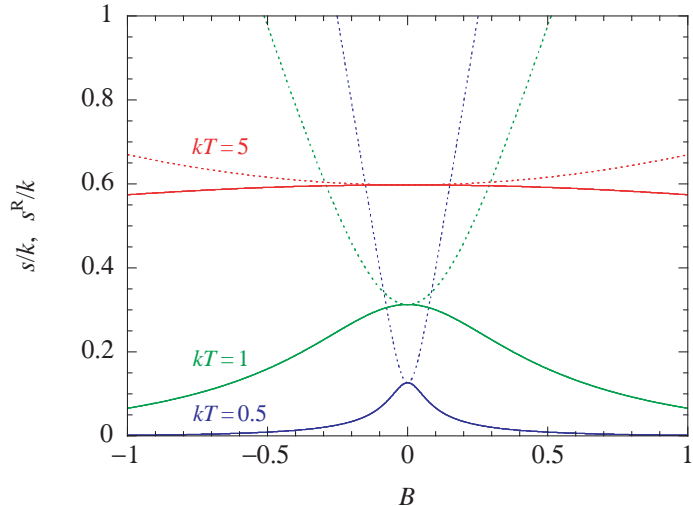}
\caption{The entropy (solid lines) and coentropy (dashed lines) per spin of the 1D Ising model for $J=0.5$ versus the external magnetic field $B$. $k$ denotes Boltzmann's constant and $T$ the temperature.  We notice that the entropy per spin tends to $s=k\ln 2$ as $kT\gg J$.}
\label{fig4}
\end{center}
\end{figure}

The free energy per spin is given by
\be
f = -kT \, \ln \Lambda_+(B) = -kT \, \ln \left(\cosh\beta B + \sqrt{\sinh^2\beta B + {\rm e}^{-4\beta J}}\right)
\ee
allowing us to obtain the standard entropy per spin by the usual thermodynamic formula
\be
s = -\left(\frac{\partial f}{\partial T}\right)_B
\ee
Now, the coentropy per spin $s^{\rm R}$ is calculated with Eq.~(\ref{coentropy}) and the average magnetization per spin (\ref{m-1DIsing}).  The entropy and the coentropy are depicted in Fig.~\ref{fig4} as a function of the external magnetic field $B$ for different values of the temperature.  Both entropies tend to increase with the temperature, as expected since they both characterize some form of disorder. If $B=0$, the coentropy is equal to the entropy. If $B\neq 0$, the coentropy is larger than the entropy because of the symmetry breaking and the general inequality (\ref{sR-s-Bm}).  For $\vert B\vert \to \infty$, the entropy decreases because of the ordering of the spins in the direction of the average magnetization, while the coentropy increases because the spin-reversed configurations become rarer and rarer.  The difference between the coentropy and the entropy behaves as
\be
s^{\rm R}-s \simeq 2 \, k \; {\rm e}^{2\beta J} \beta^2 B^2 \qquad\mbox{for}\qquad \beta B\ll 1
\ee
which explains that the separation between them is more pronounced at lower temperatures, as seen in Fig.~\ref{fig4}.

\section{The Curie-Weiss model}
\label{Curie_Weiss}

This model features a paramagnetic-ferromagnetic phase transition, which occurs at a positive critical temperature.  Below the critical temperature, a phenomenon of spontaneous symmetry breaking induces a non-vanishing magnetization in the absence of external magnetic field.  Therefore, the probability distribution of the magnetization is expected to be bimodal in finite systems, which is an interesting situation  for the fluctuation relation.

\subsection{Hamiltonian and thermodynamics}

The Hamiltonian function of the Curie-Weiss model can be written as
\be
H_N(\pmb{\sigma};B) = - \frac{J}{2N} \, M_N(\pmb{\sigma})^2  - B \, M_N(\pmb{\sigma})
\ee
in terms of the magnetization (\ref{M}) \cite{E85}.  This Hamiltonian is invariant under the whole symmetric group $G={\rm Sym}\, N$ because the Hamiltonian only depends on the total magnetization that has this invariance.  The Hamiltonian is also invariant under the group ${\mathbb Z}_2=\{1,R\}$ if $B=0$, but this symmetry is broken if $B\neq 0$.  

The total magnetization takes the $N+1$ different values $M=N-2n$ with $n=0,1,2,...,N$.
The number of spin configurations with the given magnetization $M$ is equal to
\be
C_M=\frac{N!}{\left(\frac{N+M}{2}\right)! \, \left(\frac{N-M}{2}\right)! }
\label{C_M}
\ee
In the large-system limit, this number can be approximated as
\be
C_M \simeq \sqrt{\frac{2}{\pi N (1-m^2)}} \; \exp\left[-N\left( \frac{1+m}{2}\, \ln \frac{1+m}{2} + \frac{1-m}{2}\, \ln \frac{1-m}{2}\right)\right]
\qquad\mbox{with}\quad m=\frac{M}{N}
\label{C_M-approx}
\ee
by using Stirling's formula.

Since the increment of the magnetization per spin is equal to $dm=2/N$, the partition function can thus be written as
\be
Z_N(B) = \sqrt{\frac{N}{2\pi}}\int_{-1}^{+1} dm \, \frac{{\rm e}^{-N\psi_B(m)}}{\sqrt{1-m^2}}
\ee
with the function
\be
\psi_B(m) = \frac{1+m}{2}\, \ln \frac{1+m}{2} + \frac{1-m}{2}\, \ln \frac{1-m}{2} - \frac{\beta J}{2}\, m^2 -\beta B m = \beta g(m) - \beta B m
\label{psi}
\ee
The integral can be performed with the method of steepest descents by expanding this function as
\be
\psi_B(m) = \psi_B(m_i) + \underbrace{\psi'_B(m_i)}_{=\, 0} \, (m-m_i) + \frac{1}{2}\, \psi''_B(m_i) \, (m-m_i)^2 + \cdots
\ee
around its stationary points $\{m_i\}$ such that $\psi'_B(m_i)=0$ and $\psi''_B(m_i)>0$.  The stationarity $\psi'_B(m_i)=0$ selects the value of the magnetization per spin satisfying the self-consistent condition:
\be
m_i = \tanh (\beta J m_i + \beta B)
\label{tanh}
\ee
Above the critical temperature $kT_c=J$, this equation has only one solution.  Below the critical temperature, there are three solutions, two of which satisfy the condition of local stability $\psi''_B(m_i)>0$.
In the thermodynamic limit $N\to\infty$, the partition function can thus be evaluated as
\be
Z_N(B) \simeq \sum_i \frac{{\rm e}^{-N\psi_B(m_i)}}{\sqrt{1-\beta J (1-m_i^2)}}
\label{Z-sum}
\ee
where the sum extends over the locally stable solutions with $\psi''_B(m_i)>0$.  Since the partition function is related to the free energy per spin by $Z_N(B)\simeq {\rm e}^{-N\beta f(B)}$, this latter is given by
\be
\beta f(B) = {\rm Min}_m \left\{ \psi_B(m)\right\} = \beta \, {\rm Min}_m \left\{ g(m)-B m \right\} 
\ee

Above the critical temperature, there is only one solution to Eq.~(\ref{tanh}), but two should be considered in the expression (\ref{Z-sum}) of the partition function below the critical temperature.  Since both terms are decaying exponentially, the convergence should not be expected before the size is large enough: $N\gg \vert\psi_B(m_1)-\psi_B(m_2)\vert^{-1}$.  

\begin{figure}[h]
\begin{center}
\includegraphics[scale=0.5]{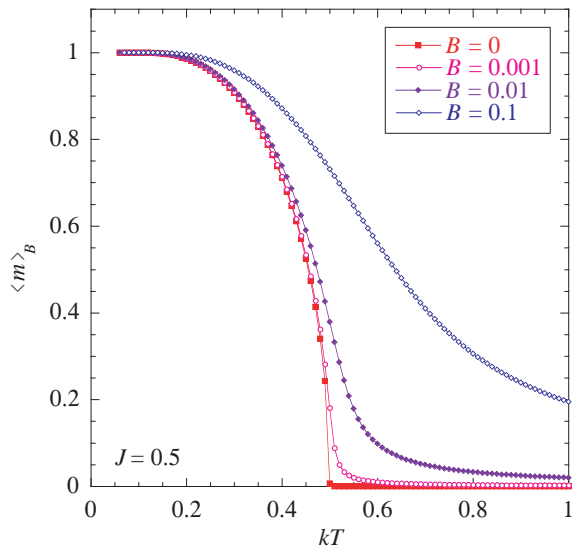}
\caption{The non-negative average total magnetization versus the thermal energy $kT$ in the Curie-Weiss model for $J=0.5$ and several values of the external magnetic field $B$.}
\label{fig5}
\end{center}
\end{figure}

Once the free energy per spin is calculated, the average magnetization per spin can be obtained by Eq.~(\ref{m_spin}), which is depicted in Fig.~\ref{fig5} for different values of the external field $B$.  The critical temperature is here equal to $kT_c=J=0.5$.  This figure allows us to appreciate the effect of the external field on the average magnetization per spin.  In order to have a moderate symmetry breaking effect, the value $B=0.001$ of the external field will be used in the following.

\subsection{Fluctuation relation}

\begin{figure}[h]
\begin{center}
\includegraphics[scale=0.5]{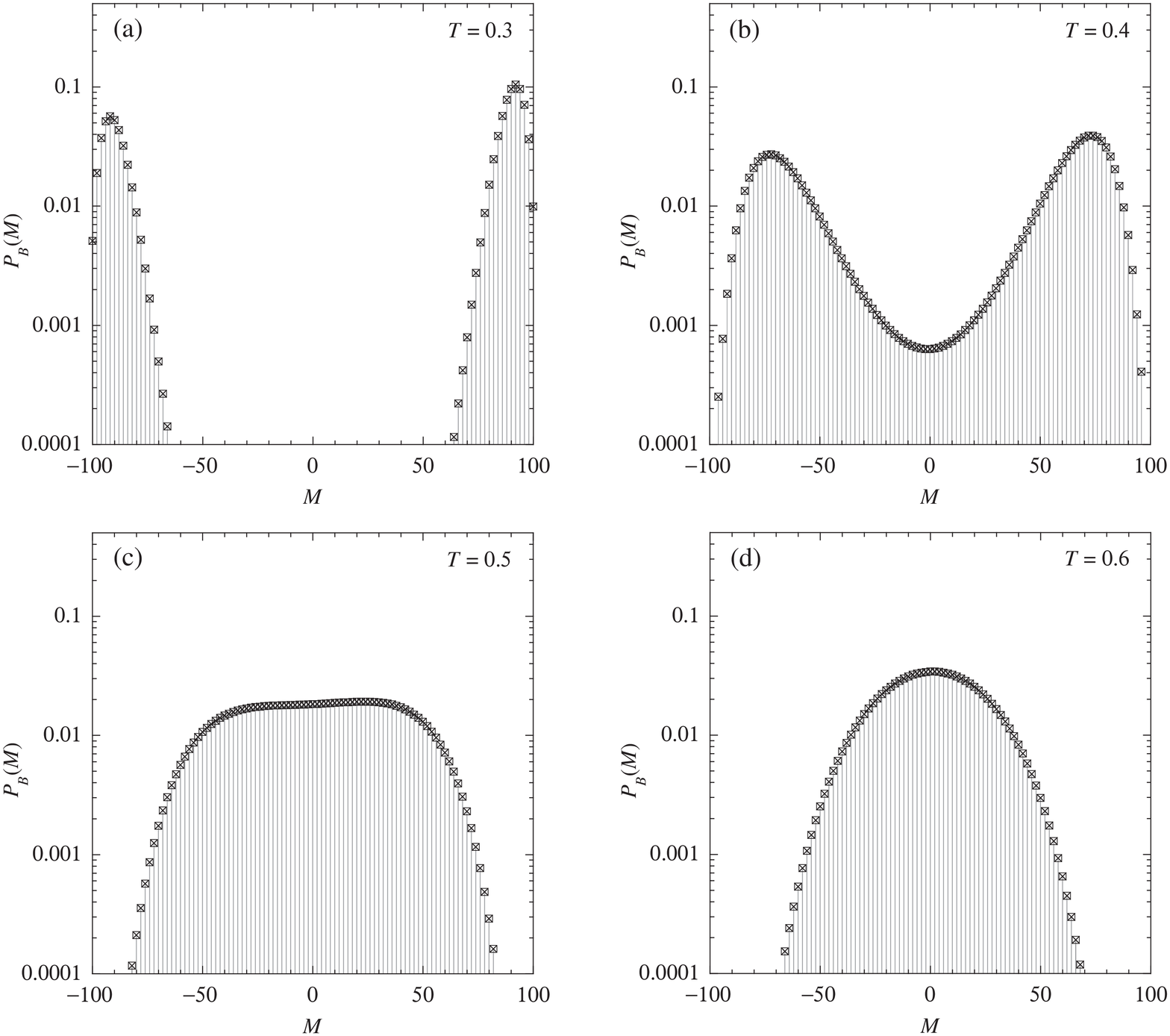}
\caption{The probability distribution $P_B(M)$ of the magnetization $M$ (open squares and lines) versus the magnetization $M$ in the Curie-Weiss model for $J=0.5$, $B=0.001$, $N=100$, and: (a)~$T=0.3$; (b)~$T=0.4$; (c)~$T=0.5$; (d)~$T=0.6$.  The crosses show the values of $P_B(-M)\exp(2\beta BM)$.  The coincidence of the crosses with the squares is the prediction of the fluctuation relation (\ref{FR}).}
\label{fig6}
\end{center}
\end{figure}

The probability distribution of the magnetization $M$ is given in the Curie-Weiss model by
\be
P_B(M) = \frac{1}{Z_N(B)}\; C_M \; \exp\left( \frac{\beta J}{2N}\, M^2 + \beta B M\right)
\label{P-CW}
\ee
where $C_M$ is the corresponding number of spin configurations (\ref{C_M}).
This distribution has been computed for $J=0.5$, $B=0.001$, $N=100$, and different values of the temperature across the phase transition, as depicted in Fig.~\ref{fig6}.  We observe that, below the critical temperature $kT_c=J=0.5$, the distribution is bimodal with two peaks centered around the two most probable values for the magnetization.  Above the transition, the distribution becomes unimodal.  The effect of the small external field $B=0.001$ is to induce an asymmetry in the distribution toward the direction of the external field.  Therefore, the average value of the magnetization has the sign of the external field.

In order to verify the fluctuation relation (\ref{FR}), the distribution $P_B(M)$ is compared in Fig.~\ref{fig6} with the prediction that it should coincide with $P_B(-M)\exp(2\beta B M)$.  We see the nice agreement between both even below the critical temperature where the distribution is bimodal and quite different from a simple Gaussian distribution.

\subsection{Large-deviation function}

For a better understanding of the behavior of the probability distribution $P_B(M)$ across the phase transition, we consider its approximation (\ref{P-Phi}) in terms of the large-deviation function (\ref{Phi}) in the thermodynamic limit $N\to\infty$.  Using the approximation (\ref{C_M-approx}), the expression (\ref{P-CW}) can be written as
\be
P_B(M) \simeq \frac{1}{Z_N(B)}\; \sqrt{\frac{2}{\pi N (1-m^2)}} \; {\rm e}^{-N\psi_B(m)}
\qquad\mbox{with}\qquad m=\frac{M}{N}
\label{P-CW-approx}
\ee
in terms of the function (\ref{psi}).  Comparing with Eq.~(\ref{P-Phi}), we recover the expression (\ref{Phi-g-f}) of the large-deviation function from Eq.~(\ref{psi}).

\begin{figure}[h]
\begin{center}
\includegraphics[scale=0.5]{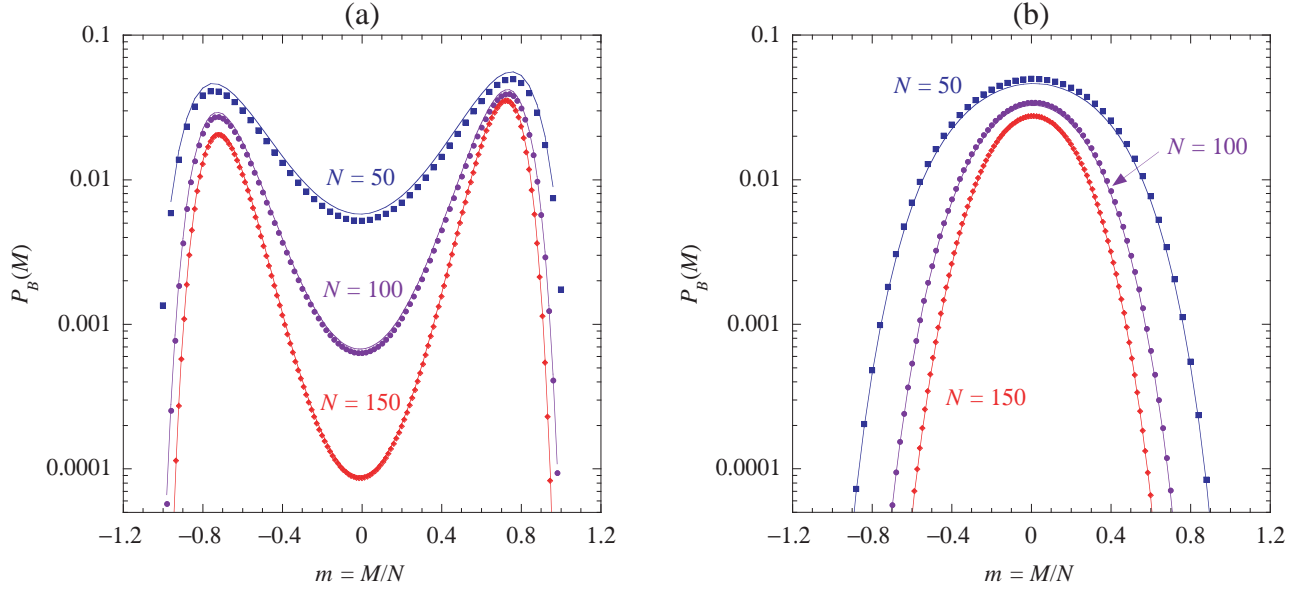}
\caption{The probability distribution $P_B(M)$ of the magnetization $M$ versus the magnetization per spin $m=M/N$ in the Curie-Weiss model for $J=0.5$, $B=0.001$, and $N=50, 100, 150$ (filled squares, circles, and diamonds) compared with its expression~(\ref{P-CW-approx}) in terms of the large-deviation function (\ref{psi}) in the large-system limit (lines): (a)~below the critical temperature at $T=0.4$ where the probability distribution is bimodal; (b)~above the critical temperature at $T=0.6$ where the probability distribution is unimodal.}
\label{fig7}
\end{center}
\end{figure}

The probability distribution (\ref{P-CW}) for $J=0.5$, $B=0.001$, and $N=50,100,150$ is compared with its large-deviation approximation (\ref{P-CW-approx}) in Fig.~\ref{fig7} below and above the critical temperature.  In this approximation, the partition function is evaluated with Eq.~(\ref{Z-sum}).  Above the critical temperature, the distribution is unimodal and a single peak contributes to the partition function.  In contrast, two peaks contribute to the partition function below the critical temperature at $kT=0.4$ in Fig.~\ref{fig7}a.  Still for $N=150$, the two terms of Eq.~(\ref{Z-sum}) are required to get the agreement seen in Fig.~\ref{fig7}a.  Indeed, the dominant peak only contributes to 63 \% of the partition function for $N=150$.  The reason is that the free energy per spin is given by $\beta f(B)=\psi_B(m_1)=-0.7308$, while the value corresponding to the subdominant peak is equal to $\psi_B(m_2)=-0.7272$.  Therefore, the second peak only becomes negligible if $N\gg \vert\psi_B(m_1)-\psi_B(m_2)\vert^{-1}\simeq 278$ for significantly larger sizes than $N=150$.  

\subsection{The cumulant generating function}

\begin{figure}[h]
\begin{center}
\includegraphics[scale=0.5]{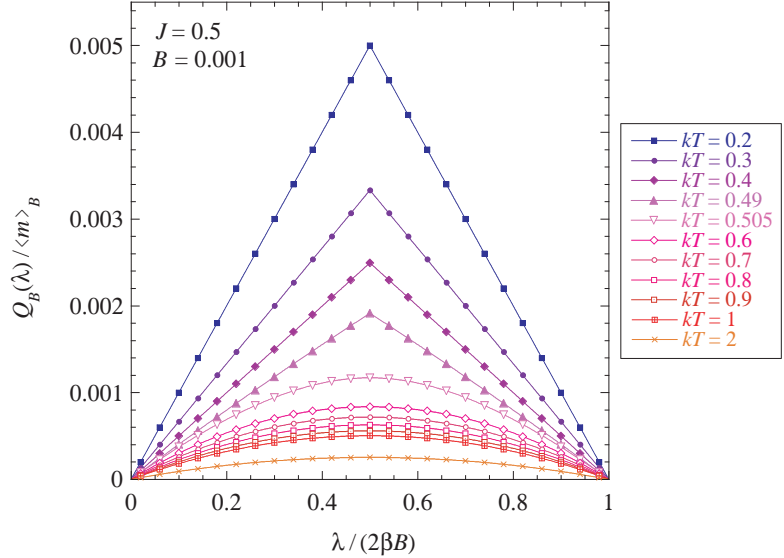}
\caption{The cumulant generating function $Q_B(\lambda)$ divided by the average magnetization per spin $\langle m \rangle_B$ versus the rescaled generating parameter $\lambda/(2\beta B)$ in the Curie-Weiss model for $J=0.5$, $B=0.001$, and different values of the temperature $T$ across the phase transition at $kT_c=J=0.5$.}
\label{fig8}
\end{center}
\end{figure}

The cumulant generating function (\ref{Q}) can be obtained using the large-deviation function (\ref{psi}).
First, the generating funciton of the statistical moments can be evaluated as
\be
\langle {\rm e}^{-\lambda M_N}\rangle_B \simeq \frac{1}{Z_N(B)} \sum_i \frac{{\rm e}^{-N\left[\psi_B(\tilde m_i)+\lambda\tilde m_i\right]}}{\sqrt{1-\beta J (1-\tilde m_i^2)}} 
\label{MGF-sum}
\ee
where the sum extends over the roots $\tilde m_i$ of $\psi'_B(\tilde m_i)+\lambda=0$ with $\psi''_B(\tilde m_i)>0$.  These roots are here the solutions of
\be
\tilde m_i = \tanh (\beta J \tilde m_i + \beta B-\lambda)
\label{tanh-lambda}
\ee
Since the partition function behaves as $Z_N(B)\simeq {\rm e}^{-N\beta f(B)}$ and $\langle {\rm e}^{-\lambda M_N}\rangle_B\sim {\rm e}^{-N Q_B(\lambda)}$ in the large-system limit, the cumulant generating function is obtained as
\be
Q_B(\lambda)={\rm Min}_m\left\{ \psi_B(m)+\lambda m -\beta f(B)\right\}={\rm Min}_m\left\{ \Phi_B(m)+\lambda m \right\}
\label{CGF-CW}
\ee
which is valid either below or above the phase transition.

For $J=0.5$ and $B=0.001$, the cumulant generating function has been calculated by Eq.~(\ref{CGF-CW}) and is shown in Fig.~\ref{fig8} for different values of the temperature across the phase transition after rescaling by the average value of the magnetization per spin.  We observe that, above the critical temperature $kT_c=J=0.5$, the cumulant generating function is differentiable where the magnetization distribution is unimodal.  In contrast, the function becomes non-differentiable below the transition where the magnetization distribution is bimodal.  The reason is the presence of two competing solutions of Eq.~(\ref{tanh-lambda}) as the parameter $\lambda$ varies.  The minimum is exchanged between these two solutions at the critical value $\lambda=\beta B$, which creates the tent-like shape of the cumulant generating function.  It should be noticed that the cumulant generating function is never exactly piecewise linear in spite of the appearance.  Its nonlinearity tends to increase with the value of the external field $B$.

The convergence is very slow near the non-differentiable point in the cumulant generating function.
By a similar reasoning as for the partition function, we should expect convergence for $N\gg\vert \psi_B(\tilde m_1)+\lambda \tilde m_1 -\psi_B(\tilde m_2)-\lambda \tilde m_2\vert^{-1}$ where $\tilde m_i=\tilde m_i(\lambda)$ are the two roots of Eq.~(\ref{tanh-lambda}).  However, this lower bound on $N$ diverges as $\vert\lambda-\beta B\vert^{-1}$ near the point of non-differentiability where the convergence is thus arbitrarily slow.

In any case, the symmetry relation (\ref{Q-sym}) is always satisfied above or below the phase transition.
We notice that, in the ferromagnetic phase, both peaks of the magnetization distribution must be taken into account in order to obtain the symmetry (\ref{Q-sym}) of the generating function.  The symmetry would not hold if only one peak would be considered.

\section{The 2D Ising model}
\label{2DIsing}

\subsection{Hamiltonian and canonical equilibrium states}

The last example studied in this paper is the famous 2D Ising model on a square lattice with nearest-neighbor interactions \cite{H87,B07}.  For finite systems, periodic boundary conditions are taken so that the square lattice forms a torus of size $L$ in both directions, containing $N=L^2$ spins one-half.  The Hamiltonian of this system is given by
\be
H_N(\pmb{\sigma};B) = \sum_{i=1}^L \sum_{j=1}^L \left[ -J \, \sigma_{i,j} \left( \sigma_{i+1,j}+\sigma_{i,j+1}\right) - B\, \sigma_{i,j}\right]
\ee
with $\sigma_{i,L+1}=\sigma_{i,1}$ and $\sigma_{L+1,j}=\sigma_{1,j}$ for $i,j=1,2,...,L$.
This Hamiltonian is invariant under the group $G$ of symmetries of the toral square lattice.  The further ${\mathbb Z}_2$ symmetry holds if $B=0$, but is broken if $B\neq 0$, as in the previous examples.

The 2D Ising model is exactly solved in the absence of external field for $B=0$ \cite{H87,B07}.  Otherwise, Monte Carlo techniques are available for computations in the presence of an external field, such as the algorithm by Metropolis~{\it et~al.} \cite{C87,LB09}, we here use.

The 2D Ising model satisfies the symmetry conditions (\ref{H-sym}) and (\ref{M-sym}) required for the fluctuation relation (\ref{FR}), as well as for the relation (\ref{D-B-M}) between the entropy and the coentropy.

\subsection{Fluctuation relation}

Here, we consider in detail the fluctuation relation (\ref{FR}) for the probability distribution of the magnetization computed by Monte Carlo simulations at different temperatures, $J=1$, and in the presence of an external magnetic field $B=0.01$ breaking the ${\mathbb Z}_2$ symmetry. We compare two lattices of finite sizes $L=10$ and $L=20$.  We notice that the paramagnetic-ferromagnetic phase transition at the critical temperature $kT_c=2.269 J$ is an exact property only in the infinite-system limit $L\to\infty$ \cite{H87}.  For systems of finite size, the transition manifests itself only as a crossover.  At low temperature, the magnetization distribution is bimodal if the external field is not too large and unimodal otherwise.

This distribution is obtained by sampling the magnetization every $10\times L^2$ Monte-Carlo spin flips generated with the algorithm by Metropolis {\it et al.} \cite{C87,LB09}.  The histogram of the magnetization is obtained with a sample of $10^7$~values for the lattice of size $L=10$ and $5\times 10^7$ values for the size $L=20$.  The histogram gives the probability density
\be
p_B(m) = \frac{1}{\Delta m} \sum_{N\left(m-\Delta m/2\right) < M < N\left(m+\Delta m/2\right)} P_B(M) 
\label{histo}
\ee
where $m=k\Delta m$ with $k=0,\pm 1,\pm 2,...,\pm K$, $\Delta m= 1/K$, and $K=L$.

\begin{figure}[h]
\begin{center}
\includegraphics[scale=0.5]{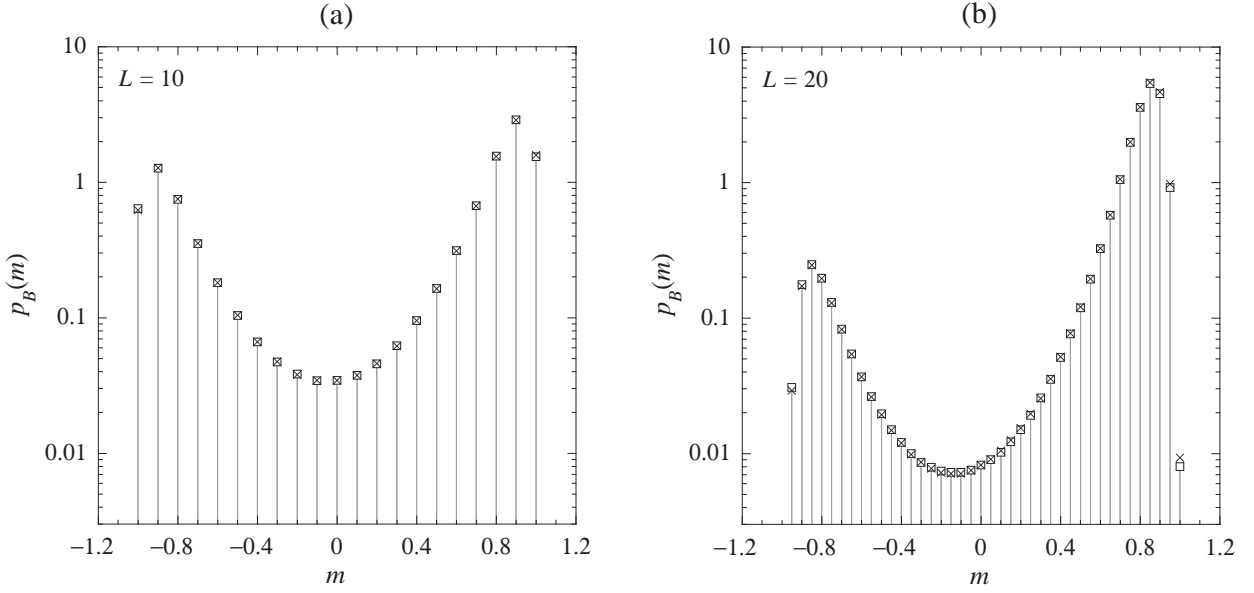}
\caption{The magnetization histogram (squares and lines) for the 2D Ising model with $J=1$, $B=0.01$, and $kT=2.2$ on a square lattice of size: (a) $L=10$; (b) $L=20$.  The crosses show the prediction of the fluctuation relation (\ref{FR-2DIsing}).}
\label{fig9}
\end{center}
\end{figure}

In Fig.~\ref{fig9}, the histogram (\ref{histo}) of the magnetization is shown for lattices of sizes $L=10$ and $L=20$ below the transition at $kT_c=2.269$.  As the size increases, we see that the bimodality of the magnetization distribution becomes stronger.

\begin{figure}[h]
\begin{center}
\includegraphics[scale=0.5]{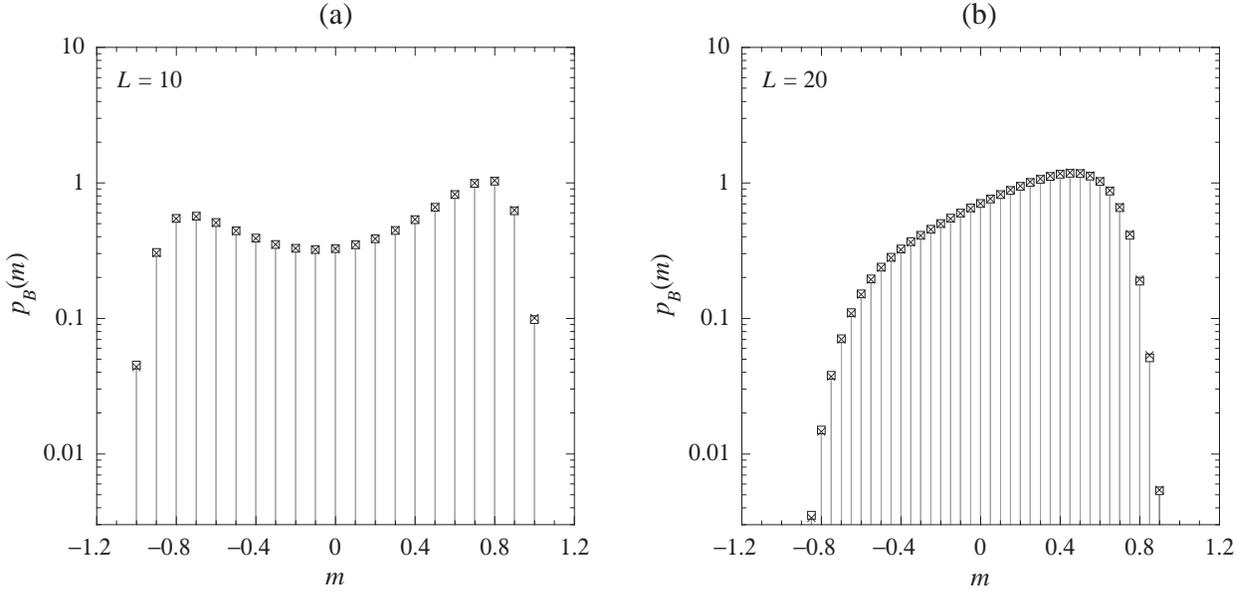}
\caption{The magnetization histogram (squares and lines) for the 2D Ising model with $J=1$, $B=0.01$, and $kT=2.5$ on a square lattice of size: (a) $L=10$; (b) $L=20$.  The crosses show the prediction of the fluctuation relation (\ref{FR-2DIsing}).}
\label{fig10}
\end{center}
\end{figure}

In Fig.~\ref{fig10}, the histogram of the magnetization is shown at $kT=2.5$ close to the transition.  We observe that the distribution is still bimodal for the small system, but it becomes unimodal as the size increases.

\begin{figure}[h]
\begin{center}
\includegraphics[scale=0.5]{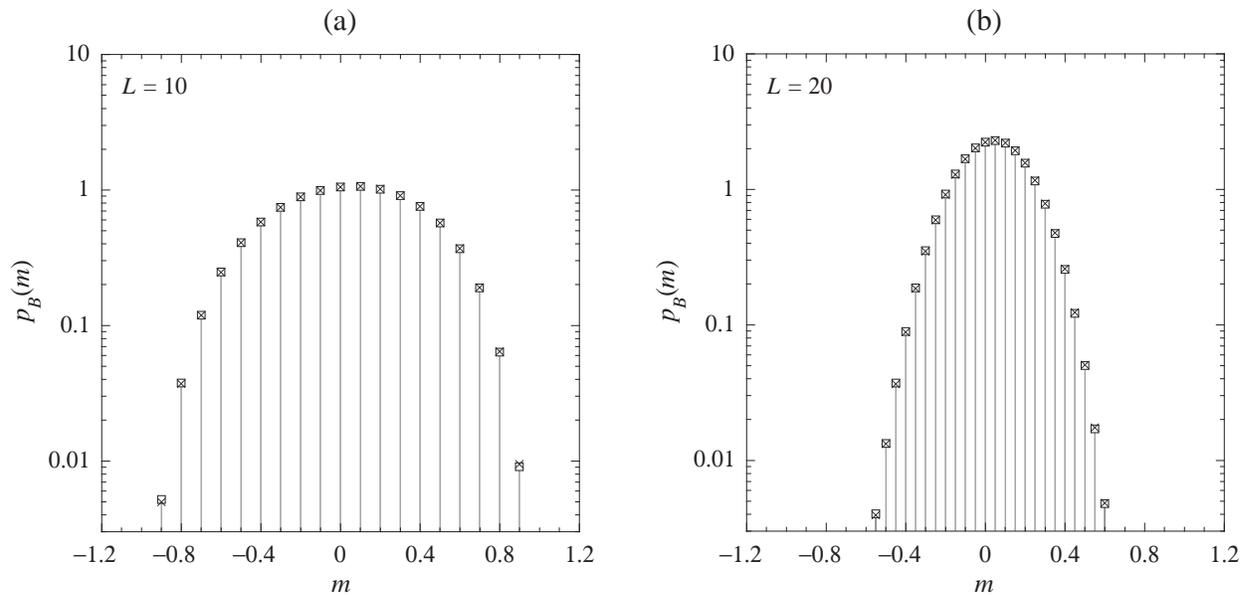}
\caption{The magnetization histogram (squares and lines) for the 2D Ising model with $J=1$, $B=0.01$, and $kT=3$ on a square lattice of size: (a) $L=10$; (b) $L=20$.  The crosses show the prediction of the fluctuation relation (\ref{FR-2DIsing}).}
\label{fig11}
\end{center}
\end{figure}

Finally, Fig.~\ref{fig11} shows the magnetization histogram at $kT=3$ above the transition where the bimodality has completely disappeared.  

Away from the critical temperature, the peaks of the magnetization distribution become sharper as the size increases, which is expected from the large-deviation expression (\ref{P-Phi}).

In all cases, the magnetization distribution computed by Monte-Carlo algorithm is compared with the prediction of the fluctuation relation (\ref{FR}) according to which
\be
p_B(m) = p_B(-m) \, \exp(2\beta B N m)
\label{FR-2DIsing}
\ee
The nice agreement observed in Figs.~\ref{fig9}-\ref{fig11} confirms the validity of the fluctuation relation in the 2D Ising model.

\section{Conclusions}
\label{Conclusions}

In this paper, relationships have been established that express the breaking of discrete symmetris by an external field in equilibrium statistical mechanics.  To be specific, we have considered spin systems described by Gibbsian canonical equilibrium states based on a Hamiltonian function of the form given by Eq.~(\ref{H}) and satisfying the symmetry conditions (\ref{H-sym})-(\ref{M-sym}) under spin reversal (\ref{spin-rev}).

With these assumptions, the exact fluctuation relation (\ref{FR}) is deduced for the probability distribution of the magnetization.  In the large-system limit, this fluctuation relation implies symmetry relations for the large-deviation function and the cumulant generating function of the fluctuating magnetization per spin.

Furthermore, a concept of coentropy or spin-reversed entropy is introduced, which forms a Kullback-Leibler divergence if combined with the standard thermodynamic entropy.  Under the assumptions considered, this Kullback-Leibler divergence is related to the product of the external field with the average magnetization.  By the non-negativity of the Kullback-Leibler divergence, this product is thus always non negative, implying a paramagnetic response to an external field.  While the entropy is a measure of disorder in typical spin configurations, the coentropy characterizes the rarity of the corresponding spin-reversed configurations.  The larger the coentropy, the rarer the spin configurations opposite to the most probable ones.

These results have been applied to four different spin models in the presence of an external magnetic field breaking the spin-reversal symmetry: the non-interacting spin model, the 1D and 2D Ising models, as well as the Curie-Weiss model.  These last two models feature a paramagnetic-ferromagnetic phase transition.  In the ferromagnetic phase, the magnetization distribution is bimodal, but the fluctuation relation is always satisfied.  In this way, the study of these models show how the established relationships characterize the breaking of spin-reversal symmetry by the external magnetic field.  In spite of the symmetry breaking phenomenon, the relationships show that the underlying fundamental symmetry (\ref{H-B}) continues to manifest itself.

\begin{table}[h]
\begin{centering}
\begin{tabular}{|c|c|c|}
\hline 
 & Equilibrium & Nonequilibrium\tabularnewline
\hline
broken symmetry & spin reversal & time reversal\tabularnewline
external control parameter & external magnetic field $B$ & affinity $A$ \tabularnewline
order parameter & average magnetization $\langle m\rangle_B$ & average current $\langle J\rangle_A$ \tabularnewline
  &  &  \tabularnewline
non-negative quantity & $\frac{1}{k}(s^{\rm R}-s)=2\beta B \langle m\rangle_B$ & entropy production $\frac{1}{k}\frac{d_{\rm i}S}{dt}=h^{\rm R}-h=A\langle J\rangle_A$ \tabularnewline
  &  &  \tabularnewline
fluctuation relation & $P_B(M) = P_B(-M)\,{\rm e}^{2\beta B M}$ & $P_A(J) = P_A(-J)\,{\rm e}^{A J}$ \tabularnewline
  &  &  \tabularnewline
\hline
\end{tabular}
\par\end{centering}
\caption{Comparison between the equilibrium and nonequilibrium broken symmetries. $k$ denotes Boltzmann's constant and $\beta=(kT)^{-1}$ the inverse temperature.}
\label{table1}
\end{table}

As shown in Table~\ref{table1}, there is an analogy with similar results already known in nonequilibrium statistical mechanics \cite{AG07JSP,G04JSP}.  In this other context, fluctuation relations have been obtained for currents flowing across open systems in nonequilibrium steady states breaking the time-reversal symmetry.  In the analogy,  the magnetization corresponds to the current and the external magnetic field to the thermodynamic force or affinity inducing non-zero values of the average current.  In this context as well, a dynamical coentropy or time-reversed entropy per unit time has been previously introduced giving the thermodynamic entropy production as a Kullback-Leibler divergence \cite{G04JSP}.  

This analogy emphasizes the role of these relationships in characterizing the breaking of some discrete ${\mathbb Z}_2$ symmetry at the statistical level of description.  In this regard, we may expect that such relationships can be used to study other discrete ${\mathbb Z}_2$ symmetries, in particular, at equilibrium.  Instead of systems with magnetic moments, systems with electric dipoles featuring transitions towards a ferroelectric phase could also be considered.  Moreover, the present relationships can be directly extended to lattice gases where spin reversal corresponds to the exchange of empty and occupied sites \cite{H87}.  Further discrete symmetries can also be envisaged such as molecular chirality in order to investigate the problem of the emergence of homochirality in physico-chemical systems, or matter-antimatter asymmetry under early cosmological conditions.  The present results also suggest to study combined broken discrete symmetries such as spin and time reversals in nonequilibrium steady states with an external magnetic field.  Beyond these contexts, similar symmetry relations have already been obtained for multifractal spectra at random critical points \cite{MBC09}.  All these advances open new perspectives for a better understanding of symmetry breaking phenomena.

\begin{acknowledgments}
Financial supports from the Francqui Foundation and the Universit\'e Libre de Bruxelles are kindly acknowledged.
\end{acknowledgments}


\end{document}